\documentclass[twocolumn]{revtex4-1}
\usepackage[dvipdfmx]{graphicx}
\usepackage{amsmath,amssymb,bm,amsthm}
\usepackage{color}
\usepackage{ulem}
\usepackage{here}
\usepackage{amsmath,amssymb,bm,amsthm,cases}
\usepackage{braket}

\def\beq{\begin{equation}}
\def\eeq{\end{equation}}
\def\nbeq{\begin{equation*}}
\def\neeq{\end{equation*}}

\def\<{\langle}
\def\>{\rangle}

\renewcommand{\d}{\partial}

\newcommand{\nc}{\newcommand}
\nc{\rnc}{\renewcommand}
\nc{\nn}{\nonumber}

\begin{document}
\title{Recurrence times of the Lieb-Liniger model in the weak and strong coupling regimes}

\author{Eriko Kaminishi$^{1}$, Takashi Mori$^2$}
\affiliation{
$^1$Quantum Computing Center, Keio University, \\3-14-1 Hiyoshi, Kohoku-ku, Yokohama, 223-8522, Japan\\
$^2$RIKEN Center for Emergent Matter Science (CEMS), Wako, 351-0198, Japan} 
\date{\today}

\begin{abstract}
Quantum systems exhibit recurrence phenomena after equilibration, but it is a difficult task to evaluate the recurrence time of a quantum system because it drastically increases as the system size increases (usually double-exponential in the number of particles) and strongly depends on the initial state.
Here, we analytically derive the recurrence times of the Lieb-Liniger model with relatively small particle numbers for the weak and strong coupling regimes.
It turns out that these recurrence times are independent of the initial state and increases only polynomially in the system size.
\end{abstract}
\maketitle

\section{Introduction}

It is of fundamental interest to study isolated quantum systems in the view of foundations of quantum statistical mechanics, i.e., equilibration and thermalization~\cite{Rigol2007, Rigol2008, Mori2018}, typicality~\cite{Tasaki1998, Goldstein2006, Reimann2008, Sugita2007}, and ergodic theorems~\cite{Neumann1929}.
It has been experimentally established that an isolated quantum system can equilibrate with respect to local (or few-body) observables, although the quantum state remains pure even after equilibration.
After equilibration, the system stays in a stationary state for a long time, but according to the quantum recurrence theorem~\cite{Ono1949, Bocchieri1957, Percival1961}, which is the quantum counterpart of Poincar\'e's recurrence theorem in classical dynamics~\cite{Poincare1890}, the system returns to a state arbitrarily close to the initial state after a sufficiently long time $T$ in the sense that $\mathcal{F}(t)=|\<\Psi(0)|\Psi(t)\>|^2\approx 1$ at $t=T$, where $|\Psi(t)\>$ is the quantum state at time $t$ and the fidelity $\mathcal{F}(t)$ is the overlap between the time-evolved state and the initial state.
In usual macroscopic quantum systems, however, the recurrence time $T$ is too long to be observed; it behaves as $e^{O(W)}$, where $W$ is the effective number of energy eigenstates contributing to the initial state.
Typically, $W$ is exponential in the number of particles $N$, and hence $T$ is double exponential in $N$.

Small quantum systems offer the possibility of investigating full relaxation dynamics of isolated quantum systems, including the recurrence.
It has been recognized that a relatively small quantum system such as the one with $N\sim 10$ atoms can equilibrate because of a huge dimensionality of the Hilbert space.
Moreover, such small systems may exhibit recurrence long after equilibration but within an experimentally accessible timescale.
Indeed, recurrence behavior has been observed in the Jaynes-Cummings system~\cite{Rempe1987}, in ultracold atoms in an optical lattice~\cite{Greiner2002, Will2010} as well as a large system with thousands of atoms in one-dimensional box-shaped potential by focusing on specific observables that are related to the low-energy collective motion~\cite{Rauer2018}.
It should be noted that recurrence after equilibration (or ``collapse and revival'') in a small system is a purely quantum phenomenon which has no classical counterpart (see Ref.~\cite{Mori2018}).
Since compatibility of equilibration and recurrence is a heart of the emergence of irreversibility from reversible microscopic dynamics, the long-term dynamics of small isolated quantum systems is of fundamental interest.

Theoretically, in Ref.~\cite{Kaminishi2015}, it has been numerically reported that the Lieb-Liniger model~\cite{Lieb1963}, which is a representative model of a one-dimensional Bose gas, exhibits recurrence phenomena within an experimentally accessible timescale; the recurrence time increases only polynomially with $N$.
However, numerical results in Ref.~\cite{Kaminishi2015} have been obtained only for a special initial state (a uniform superposition of yrast states), and it has been unclear whether those polynomial dependences of the recurrence time are generic features of the Lieb-Liniger model.

In this paper, we evaluate the recurrence time in the weak and the strong coupling regimes.
It is shown that the recurrence time is polynomial in the system size in both regimes.
Our finding implies that no fine tuning of the initial state is necessary to see recurrence phenomena in a small one-dimensional Bose gas.
The remaining part of the paper is organized as follows.
In Sec.~\ref{sec:recurrence}, we introduce the model and explain the definition of the recurrence.
In Sec.~\ref{sec:weak}, we show that the recurrence time is proportional to $L/c$, see eq.~(\ref{eq:rec_weak}), for generic initial states in the weak-coupling regime.
In Sec.~\ref{sec:strong}, we show that the recurrence time is roughly proportional to $L^3c$, see eq.~(\ref{eq:rec_strong}), for generic initial states in the strong-coupling regime.
These results are general, but $T$ may be further shortened by an accident for some special initial states.
Such an accidental example is provided in Sec.~\ref{sec:dark}.

\section{Recurrence time in the Lieb-Liniger model}
\label{sec:recurrence}

The Lieb-Liniger (LL) model is a prototypical integrable model of the one-dimensional Bose gas.
Because of its integrability and experimental relevance, the relaxation dynamics of the Lieb-Liniger model has been actively studied~\cite{Caux2007, Mossel2012, Deepak2012, Sato2012, Girardeau2003, Kaminishi2015, Zill2015, Sato2016, Kaminishi2018}.
The Hamiltonian of the LL model is given by
\beq
H=-\sum_{i=1}^N\frac{\d^2}{\d q_i^2}+2c\sum_{i<j}\delta(q_i-q_j),
\eeq
where $q_i$ is the coordinate of $i$th particle that satisfies $0\leq q_i\leq L$.
We consider the repulsive interactions $c>0$ and assume the periodic boundary condition.
Energy eigenstates are characterized by a set of Bethe quantum numbers $\bm{I}^N\equiv\{ I_1,I_2,\dots,I_N\}$ with $I_1<I_2<\dots<I_N$.
When the number of particles is odd, $I_j$ is integer, and otherwise $I_j$ is half-odd integer.
When the set of Bethe quantum numbers is fixed, the set of Bethe momenta $\bm{k}^N\equiv\{k_1,k_2,\dots,k_N\}$ is determined by solving the Bethe ansatz equation:
\beq
k_j=\frac{2\pi}{L}I_j-\frac{2}{L}\sum_{l(\neq j)}^N\arctan\left(\frac{k_j-k_l}{c}\right).
\eeq
The total momentum $P$ and the total energy $E$ are then given by
\beq
P(\bm{k}^N)=\sum_{j=1}^Nk_j, \qquad
E(\bm{k}^N)=\sum_{j=1}^Nk_j^2,
\eeq
respectively.
Corresponding to $I_1<I_2<\dots<I_N$, $k_1<k_2<\dots<k_N$ holds.
When the initial state is expanded as
\beq
|\Psi(0)\>=\sum_{\bm{k}^N}C_{\bm{k}^N}|\bm{k}^N\>,
\label{eq:initial}
\eeq
the state at time $t$ is given by
\beq
|\Psi(t)\>=\sum_{\bm{k}^N}C_{\bm{k}^N}e^{-iE(\bm{k}^N)t}|\bm{k}^N\>.
\eeq
In this paper, the recurrence time $T$ is defined by the shortest time at which the fidelity
\beq
\mathcal{F}(t)=|\<\psi(0)|\psi(t)\>|^2=\left|\sum_{\bm{k}^N}|C_{\bm{k}^N}|^2e^{-iE(\bm{k}^N)t}\right|^2
\eeq
exceeds some threshold after the first decay of $\mathcal{F}(t)$.
Here we set the threshold as 0.9, so we say that the recurrence occurs at $t=T$ when
\beq
\mathcal{F}(T)\geq 0.9.
\eeq
At the recurrence time $T$, all the phases $E(\bm{k}^N)T$ are almost identical modulo $2\pi$, that is, $E(\bm{k}^N)T\simeq \theta+2\pi m(\bm{k}^N)$ with $\theta\in[0,2\pi)$ and $\{ m(\bm{k}^N)\}$ integers.

\section{Weak-coupling regime}
\label{sec:weak}

\subsection{Weak-coupling expansion}
The recurrence time is determined by the structure of the spectrum $\{ E(\bm{k}^N)\}$ and the type of the initial state.
Therefore, we should investigate the Bethe ansatz equation in order to evaluate the recurrence time.

First, we consider the weak-coupling regime, $c\ll 1$.
In that case,
\beq
\arctan\left(\frac{k_j-k_l}{c}\right)\simeq \frac{\pi}{2}{\rm sgn}(k_j-k_l)-\frac{c}{k_j-k_l}.
\eeq
Since $k_j<k_l$ if and only if $j<l$, ${\rm sgn}(k_j-k_l)={\rm sgn}(j-l)$.
The Bethe ansatz equation becomes
\beq
k_j=\frac{2\pi}{L}I_j-\frac{\pi}{L}\sum_{l(\neq j)}^N{\rm sgn}(j-l)+\frac{2c}{L}\sum_{l(\neq j)}\frac{1}{k_j-k_l}.
\eeq
Here,
\beq
\sum_{l(\neq j)}^N{\rm sgn}(j-l)=\sum_{l=1}^{j-1}(+1)+\sum_{l=j+1}^N(-1)=2j-1-N,
\eeq
so we obtain
\beq
k_j=\frac{2\pi}{L}\left[I_j-\left(j-\frac{N+1}{2}\right)\right]+\frac{2c}{L}\sum_{l(\neq j)}\frac{1}{k_j-k_l}.
\eeq
It is noted that $I_j^{\rm G}=j-(N+1)/2$ is the Bethe quantum number corresponding to the ground state.
Thus we obtain the Bethe ansatz equation for the weak-coupling regime:
\beq
k_j=\frac{2\pi}{L}(I_j-I_j^{\rm G})+\frac{2c}{L}\sum_{l(\neq j)}\frac{1}{k_j-k_l}.
\label{eq:Bethe_weak}
\eeq

In order to solve the above equation, we expand the Bethe momentum $k_j$ as
\beq
k_j=k_j^{(0)}+c^{1/2}k_j^{(1/2)}+ck_j^{(1)}+o(c).
\label{eq:k_weak}
\eeq
The term proportional to $c^{1/2}$ is necessary to satisfy eq.~(\ref{eq:Bethe_weak}) ~\cite{Gaudin2014}.
By substituting eq.~(\ref{eq:k_weak}) to eq.~(\ref{eq:Bethe_weak}), we obtain
\begin{align}
&k_j^{(0)}=\frac{2\pi}{L}(I_j-I_j^{\rm G}),
\label{eq:weak_k0}\\
&k_j^{(1/2)}=\frac{2}{L}\sum_{\substack{l(\neq j) \\ (k_l^{(0)}=k_j^{(0)})}}\frac{1}{k_j^{(1/2)}-k_l^{(1/2)}}, 
\label{eq:weak_k1/2}\\
&k_j^{(1)}=\frac{2}{L}\sum_{\substack{l \\ (k_l^{(0)}\neq k_j^{(0)})}}\frac{1}{k_j^{(0)}-k_l^{(0)}}.
\label{eq:weak_k1}
\end{align}

$k_l^{(0)}=k_j^{(0)}$ implies there is no hole between $I_l$ and $I_j$, while $k_l\neq k_j$ implies there is at least one hole between $I_l$ and $I_j$.
Let us divide the set of Bethe quantum numbers into some groups by the following rule:
$$I_{j+1}=I_j+1 \iff \text{ $j$ and $j+1$ belong to the same group.}$$
In other words,
\begin{align*}
&k_j^{(0)}=k_l^{(0)} \iff \text{$j$ and $l$ belong to the same group, and} \\
&k_j^{(0)}\neq k_l^{(0)} \iff \text{$j$ and $l$ belong to different groups.}
\end{align*}
Each group is denoted by $\mathcal{G}_a$ and the number of elements of $\mathcal{G}_a$ is denoted by $n_a$.
If $i$ and $j$ belong to the same group, we write $i\sim j$, and if $i$ and $j$ belong to the different groups, we write $i\nsim j$.
The Bethe momentum of the group $\mathcal{G}_a$ is denoted by $k^{(0)}(a)$.
Note that $k_i^{(0)}=k^{(0)}(a)$ if and only if $i\in\mathcal{G}_a$.

\subsection{Properties of $k_j^{(1/2)}$ and $k_j^{(1)}$}

Equation~(\ref{eq:weak_k1/2}) cannot be solved, but some useful formulas can be derived.
Due to the antisymmetry, we have
\beq
\sum_{j\in\mathcal{G}_a}k_j^{(1/2)}=0.
\label{eq:property_1/2-1}
\eeq
Moreover, we can prove the convenient relation
\beq
\sum_{j\in\mathcal{G}_a}\left(k_j^{(1/2)}\right)^2=\frac{n_a(n_a-1)}{L}.
\label{eq:property_1/2-2}
\eeq
Equation (\ref{eq:property_1/2-2}) is proven in the following way.
The Bethe momentum corrections $\{k_j^{(1/2)}\}$ in the same group $\mathcal{G}_a$ satisfy eq.~(\ref{eq:weak_k1/2}), which is written in the form
\beq
x_j=\sum_{l(\neq j)}^M\frac{1}{x_j-x_l}
\eeq
for $j=1,2,\dots,M$.
Here, $x$ corresponds to $k\sqrt{L/2}$ and $M$ corresponds to $n_a$.
Assuming $j\neq M$, we decompose the summation over $l$ into $l\leq M-1$ and $l=M$,
\beq
x_j=\sum_{l(\neq j)}^{M-1}\frac{1}{x_j-x_l}+\frac{1}{x_j-x_M}.
\eeq
From this equation we obtain
\beq
x_j^2-x_jx_M-1=\sum_{l(\neq j)}^{M-1}\frac{1}{x_j-x_l}.
\eeq
By taking the summation over $j\leq M-1$, we have
\begin{align}
\sum_{j=1}^{M-1}x_j^2-x_M\sum_{j=1}^{M-1}x_j-(M-1)
\nonumber \\
=\sum_{j\neq l}^{M-1}\frac{x_j}{x_j-x_l} -x_M\sum_{j\neq l}^{M-1}\frac{1}{x_j-x_l}.
\label{eq:xj_sum}
\end{align}
Equation~(\ref{eq:property_1/2-1}) implies $\sum_{j=1}^Mx_j=0$ and therefore $\sum_{j=1}^{M-1}x_j=-x_M$.
In addition, the last term of the right-hand-side of eq.~(\ref{eq:xj_sum}) is zero due to anti-symmetry.
Thus we obtain
\beq
\sum_{j=1}^Mx_j^2=M-1+\sum_{j\neq l}^{M-1}\frac{x_j}{x_j-x_l}.
\label{eq:xj_square}
\eeq
By putting $X=\sum_{j\neq l}^{M-1}x_j/(x_j-x_l)$,
\beq
X=\sum_{j\neq l}^{M-1}\left(1+\frac{x_l}{x_j-x_l}\right)
=(M-1)(M-2)-X.
\eeq
Thus $X=(M-1)(M-2)/2$.
By substituting it into eq.~(\ref{eq:xj_square}), we obtain
\beq
\sum_{j=1}^Mx_j^2=\frac{M(M-1)}{2},
\eeq
which is equivalent to eq.~(\ref{eq:property_1/2-2}).

From eq.~(\ref{eq:weak_k1}),
\begin{align}
k_j^{(1)}&=\frac{2}{L}\sum_{l(\nsim j)}\frac{1}{k_j^{(0)}-k_l^{(0)}}
\nonumber \\
&=\frac{2}{L}\sum_{\substack{b \\ (j\notin\mathcal{G}_b)}}\frac{n_b}{k_j^{(0)}-k^{(0)}(b)}.
\end{align}
From this expression, we have
\beq
\sum_{j\in\mathcal{G}_a}k_j^{(1)}=\frac{2}{L}\sum_{b(\neq a)}\frac{n_an_b}{k^{(0)}(a)-k^{(0)}(b)}.
\label{eq:property_1}
\eeq

\subsection{Approximate energy spectrum}

The energy in the weak-coupling regime is given by
\begin{align}
E(\bm{k}^N)&=\sum_{j=1}^Nk_j^2
\nonumber \\
&\simeq\sum_{j=1}^N\left[k_j^{(0)}+c^{1/2}k_j^{(1/2)}+ck_j^{(1)}\right]^2
\nonumber \\
&=\sum_a\sum_{j\in\mathcal{G}_a}\left[k^{(0)}(a)+c^{1/2}k_j^{(1/2)}+ck_j^{(1)}\right]^2
\nonumber \\
&=\sum_a\left\{n_ak^{(0)}(a)^2+2c^{1/2}k^{(0)}(a)\sum_{j\in\mathcal{G}_a}k_j^{(1/2)}\right.
\nonumber \\
&\quad \left.  +c\sum_{j\in\mathcal{G}_a}\left(k_j^{(1/2)}\right)^2+2ck^{(0)}(a)\sum_{j\in\mathcal{G}_a}k_j^{(1)}\right\}+\mathcal{O}(c^2).
\end{align}
By using eqs.~(\ref{eq:property_1/2-1}), (\ref{eq:property_1/2-2}), and (\ref{eq:property_1}), we obtain
\begin{align}
E(\bm{k}^N)&=\sum_an_ak^{(0)}(a)^2+\frac{c}{L}\sum_an_a(n_a-1)
\nonumber \\
&\quad +\frac{4c}{L}\sum_{a\neq b}n_an_b\frac{k^{(0)}(a)}{k^{(0)}(a)-k^{(0)}(b)}+\mathcal{O}(c^2).
\end{align}

Putting 
\beq
A=\sum_{a\neq b}n_an_b\frac{k^{(0)}(a)}{k^{(0)}(a)-k^{(0)}(b)},
\eeq
we have
\begin{align}
A&=\sum_{a\neq b}n_an_b+\sum_{a\neq b}n_an_b\frac{k^{(0)}(b)}{k^{(0)}(a)-k^{(0)}(b)}
\nonumber \\
&=\sum_an_a\sum_bn_b-\sum_an_a^2-A
\nonumber \\
&=N^2-\sum_an_a^2-A.
\end{align}
Solving this equation, we obtain
\beq
A=\frac{1}{2}\left(N^2-\sum_an_a^2\right),
\eeq
which is integer or half-integer.
Thus we obtain the approximate energy spectrum for the weak-coupling regime,
\begin{align}
E(\bm{k}^N)=E_w^{(0)}(\bm{k}^N)+V_w(\bm{k}^N)+E_w'(\bm{k}^N).
\label{eq:E_w}
\end{align}
The first term $E_w^{(0)}(\bm{k}^N)=\sum_an_ak^{(0)}(a)^2$ is the energy without interactions.
The second term
\beq
V_w(\bm{k}^N)=\frac{c}{L}\left[N(2N-1)-\sum_an_a^2\right]
\label{eq:V_w}
\eeq
represents the weak-coupling correction.
The last term $E_w'(\bm{k}^N)=\mathcal{O}(c^2)$ is the higher-order contribution, which we assume to be negligible.
It is noted that the weak-coupling correction (\ref{eq:V_w}) is an integer multiple of $c/L$.
This characteristic of the energy spectrum makes the recurrence time drastically short compared to that in a generic quantum system.

\subsection{Evaluation of the recurrence time}

The recurrence occurs at time $T$ if $E(\bm{k}^N)T\simeq E_w^{(0)}(\bm{k}^N)T+V_w(\bm{k}^N)T$ are almost identical modulo $2\pi$ for all $\bm{k}^N$ with $C_{\bm{k}^N}\neq 0$.
Since the timescales of $E_w^{(0)}$ and $V$ are separated for small $c$, as long as no special reason is present (but see sec.~\ref{sec:dark}), each of $E_w^{(0)}(\bm{k}^N)T$ and $V_w(\bm{k}^N)T$ would be independent of $\bm{k}^N$ (mod $2\pi$), that is, 
\begin{align}
E_w^{(0)}(\bm{k}^N)T&\approx \theta+2\pi m(\bm{k}^N) 
\label{eq:rec0}\\
V_w(\bm{k}^N)T&\approx \theta'+2\pi m'(\bm{k}^N),
\label{eq:rec1}
\end{align}
with integers $m(\bm{k}^N)$ and $m'(\bm{k}^N)$.

Equation~(\ref{eq:rec0}) implies
\beq
T\sum_an_ak^{(0)}(a)^2\approx \theta+2\pi m(\bm{k}^N).
\eeq
Because $\sum_an_ak^{(0)}(a)^2=(2\pi /L)^2\times (\rm integer)$, we have
\beq
T\simeq\frac{L^2}{2\pi}\times ({\rm integer})\equiv T_w^{(0)}\times (\rm integer),
\label{eq:T0}
\eeq
where $T_w^{(0)}=L^2/2\pi$ is the recurrence time in the ideal Bose gas in one dimension~\cite{Kaminishi2015}.

Next, $T$ must also satisfy eq.~(\ref{eq:rec1}), that is,
\beq
T\frac{c}{L}\sum_an_a^2\simeq \theta'+2\pi m'(\bm{k}^N).
\eeq
Here, we notice that
\beq
N^2-\sum_an_a^2=2\sum_{a<b}n_an_b
\eeq
is even.
It implies that the shortest time at which eq.~(\ref{eq:rec1}) is fulfilled is given by
\beq
T\simeq\frac{\pi L}{c}\equiv T_w.
\label{eq:Tw}
\eeq
Since the initial state is a superposition of several eigenstates as eq.~(\ref{eq:initial}), the perturbation energy $V_w(\bm{k}^N)$ has some fluctuations.
If this uncertainty of the perturbation energy is denoted by
\beq
\Delta V_w=\left(\<V_w^2\>_{\Psi}-\<V_w\>_{\Psi}^2\right)^{1/2},
\eeq
where $\<F\>_{\Psi}=\sum_{\bm{k}^N}|C_{\bm{k}^N}|^2F(\bm{k}^N)$ is the average in the initial state, eq.~(\ref{eq:rec1}) approximately holds during the duration time $\Delta T\sim 2\pi/\Delta V_w$ around $t=T_w$.
When $\Delta T$ is greater than $T_w^{(0)}$, i.e.,
\beq
\frac{1}{2\pi}T_w^{(0)}\Delta V_w=\frac{L^2}{(2\pi)^2}\Delta V_w\lesssim 1
\label{eq:condition}
\eeq
is satisfied, equations~(\ref{eq:T0}) and (\ref{eq:Tw}) can simultaneously hold, and hence the recurrence occurs at $t=T_w=\pi L/c$.

Up to here, we have neglected higher-order contribution $E_w'(\bm{k}^N)$ in eq.~(\ref{eq:E_w}).
Another condition of the recurrence at $T_w$ is that this higher-order contribution is actually negligible up to $t=T_w$.
The phase factor at time $t=T_w$ due to the higher-order contribution is given by $E_w'(\bm{k}^N)T_w$, which should be almost independent of $\bm{k}^N$.
That is,
\beq
T_w\Delta E_w'\lesssim 1
\label{eq:condition2}
\eeq
should be satisfied, where $\Delta E_w'$ is the quantum fluctuation of $E_w'(\bm{k}^N)$:
\beq
\Delta E_w'=\left(\<(E_w')^2\>_\Psi-\<E_w'\>_\Psi^2\right)^{1/2}.
\eeq

In summary, when the conditions (\ref{eq:condition}) and (\ref{eq:condition2}) are fulfilled, we expect that the recurrence time $T$ is given by
\beq
T\simeq T_w=\frac{\pi L}{c}.
\label{eq:rec_weak}
\eeq
The conditions (\ref{eq:condition}) and (\ref{eq:condition2}) restrict the system size.
When the system size is too large, these conditions are not satisfied and the recurrence would not happen.
The critical value of the system size below which the recurrence occurs depends on the initial state and is not universal since $\Delta V_w$ and $\Delta E_w'$ depend on the initial state.
On the other hand, apart from these conditions, we have made no assumption on the initial state.
In the weak-coupling regime, the recurrence occurs around at $T_w$ given by eq.~(\ref{eq:rec_weak}) \textit{for any initial state}.

\section{Strong-coupling regime}
\label{sec:strong}

Let us consider the strong-coupling regime.
When $1/c\ll 1$,
\beq
\arctan\left(\frac{k_j-k_l}{c}\right)\approx\frac{1}{c}(k_j-k_l).
\eeq
The Bethe ansatz equation reads
\beq
k_j\simeq\frac{2\pi}{L}I_j-\frac{2}{L}\sum_{l(\neq j)}\frac{k_j-k_l}{c}.
\eeq
This equation is solved as
\beq
k_j=\frac{1}{1+\frac{2\rho}{c}}\frac{2\pi}{L}\left(I_j+\frac{2}{Lc}I_\mathrm{tot}^{(1)}\right),
\eeq
where $\rho=N/L$ is the number density and we use the following notation:
\beq
I_{\rm tot}^{(1)}=\sum_{j=1}^NI_j, \qquad
I_{\rm tot}^{(2)}=\sum_{j=1}^NI_j^2.
\eeq
the energy spectrum is given by
\begin{align}
E(\bm{k}^N)&=\sum_{j=1}^Nk_j^2
\nonumber \\
&=E_s^{(0)}(\bm{k}^N)+V_s(\bm{k}^N)+E_s'(\bm{k}^N),
\end{align}
where
\beq
E_s^{(0)}=\left(\frac{2\pi}{L}\right)^2I_\mathrm{tot}^{(2)}
\eeq
is the energy without interactions,
\beq
V_s=-\frac{16\pi^2}{L^3c}\left(1-\frac{\rho}{3c}\right)\left[NI_{\rm tot}^{(2)}-\left(I_{\rm tot}^{(1)}\right)^2\right]
\eeq
is the correction term up to $1/c^2$, and $E_s'(\bm{k}^N)=\mathcal{O}(1/c^3)$ is the higher-order contribution, which is assumed to be negligible.

The recurrence time in the strong-coupling regime is evaluated by imposing the conditions
\begin{align}
E_s^{(0)}(\bm{k}^N)T&\simeq\theta+2\pi m(\bm{k}^N) 
\label{eq:rec0_s},\\
V_s(\bm{k}^N)T&\simeq\theta'+2\pi m'(\bm{k}^N)
\label{eq:rec1_s}
\end{align}
with some integers $m(\bm{k}^N)$ and $m'(\bm{k}^N)$ if there is no special reason.

Equation~(\ref{eq:rec0_s}) implies
\beq
T=T_s^{(0)}:=\left\{
\begin{aligned}
\frac{L^2}{2\pi}\times({\rm integer}) \qquad \text{for odd $N$}, \\
\frac{L^2}{4\pi}\times({\rm integer}) \qquad \text{for even $N$},
\end{aligned}
\right.
\eeq
which is the recurrence time in the Tonks-Girardeau gas ($c=+\infty$)~\cite{Girardeau960relationship}.
The reason why the recurrence time for even $N$ is half of that for odd $N$ is the following.
For even $N$, $I_j=(2n_j+1)/2$ with integer $n_j$, and in that case
\beq
I_{\rm tot}^{(2)}=\sum_{j=1}^Nn_j(n_j+1)+{\rm const.}
\eeq
The first term is always even, which halves the recurrence time.

The recurrence time $T$ must also satisfy eq.~(\ref{eq:rec1_s}).
Since $NI_{\rm tot}^{(2)}-(I_{\rm tot}^{(1)})^2$ is always integer,
\beq
T\simeq\frac{L^3c}{8\pi\left(1-\frac{\rho}{3c}\right)}=:T_s^{(\rm even)}
\label{eq:rec_strong}
\eeq
satisfies eq.~(\ref{eq:rec1_s}).
When $N$ is odd, $I_{\rm tot}^{(2)}=\sum_{j=1}^NI_j(I_j-1)+I_{\rm tot}^{(1)}$ and each $I_j$ is an integer, which implies that $I_{\rm tot}^{(2)}$ is even (odd) when $I_{\rm tot}^{(1)}$ is even (odd), respectively.
Therefore, $NI_{\rm tot}^{(2)}-I_{\rm tot}^{(1)2}$ is always even, which halves the recurrence time for odd $N$:
\beq
T\simeq\frac{L^3c}{16\pi\left(1-\frac{\rho}{3c}\right)}=:T_s^{(\rm odd)}.
\eeq
We define $T_s$ as
\beq
T_s:=\begin{cases}
T_s^{(\rm even)} &\text{when $N$ is even},\\
T_s^{(\rm odd)} &\text{when $N$ is odd}.
\end{cases}
\eeq

In order that the recurrence actually occurs at $t\simeq T_s$, the duration time of the recurrence $\Delta T$ must be greater than $T_s^{(0)}$.
Since the duration of the recurrence is roughly given by $\Delta T\sim 2\pi/\Delta V_s$, where $\Delta V_s=(\<V_s^2\>_{\Psi}-\<V_s\>_{\Psi}^2)^{1/2}$ is the fluctuation of the interaction energy, it implies
\beq
\frac{1}{2\pi}T_s^{(0)}\Delta V_s\lesssim 1.
\label{eq:condition_s}
\eeq

The condition that the contribution from higher-order terms $E_s'(\bm{k}^N)$ is actually negligible up to the time $t=T_s$ is given by
\beq
T_s\Delta E_s'\lesssim 1,
\label{eq:condition2_s}
\eeq
where $\Delta E_s'=(\<(E_s')^2\>-\<E_s\>^2)^{1/2}$ is the fluctuation of the higher-order contribution to the energy.

Again, the conditions (\ref{eq:condition_s}) and (\ref{eq:condition2_s}) restrict the system size at which the recurrence is observed.
Apart from these restrictions, we do not assume the specific initial state to derive Eq.~(\ref{eq:rec_strong}), and thus the recurrence occurs at $t=T_s$ for an arbitrary initial state in the strong coupling regime.

\section{Numerical demonstration of recurrence phenomena}
\label{sec:numerical}

We numerically demonstrate the recurrence in the weak and the strong coupling regimes.
We construct the initial state as
\beq
|\Psi(0)\rangle={\sum_{\bm{k}^N}}'C_{\bm{k}^N}|\bm{k}^N\rangle,
\label{eq:initial_num}
\eeq
where the prime implies that the summation is restricted to the Bethe eigenstates whose Bethe quantum numbers $\{I_j\}$ satisfy
\beq
I_1^\mathrm{G}-2\leq I_1<I_2<\dots<I_N\leq I_N^\mathrm{G}+2.
\eeq
Expansion coefficients in eq.~(\ref{eq:initial}) are chosen as 
\beq
C_{\bm{k}^N}=\frac{e^{-\beta E(\bm{k}^N)/2}}{\sqrt{Z}}
\eeq
with $Z=\sum_{\bm{k}^N}'e^{-\beta E(\bm{k}^N)}$.
By controlling the value of $\beta$, we can prepare several initial states for fixed $L$, $N$, and $c$.
Without loss of generality, we put $L=N$.

\begin{figure*}[tb]
\begin{minipage}{0.33\hsize}
\begin{tabular}{c}
(a) $N=8$\\
\includegraphics[width=5cm]{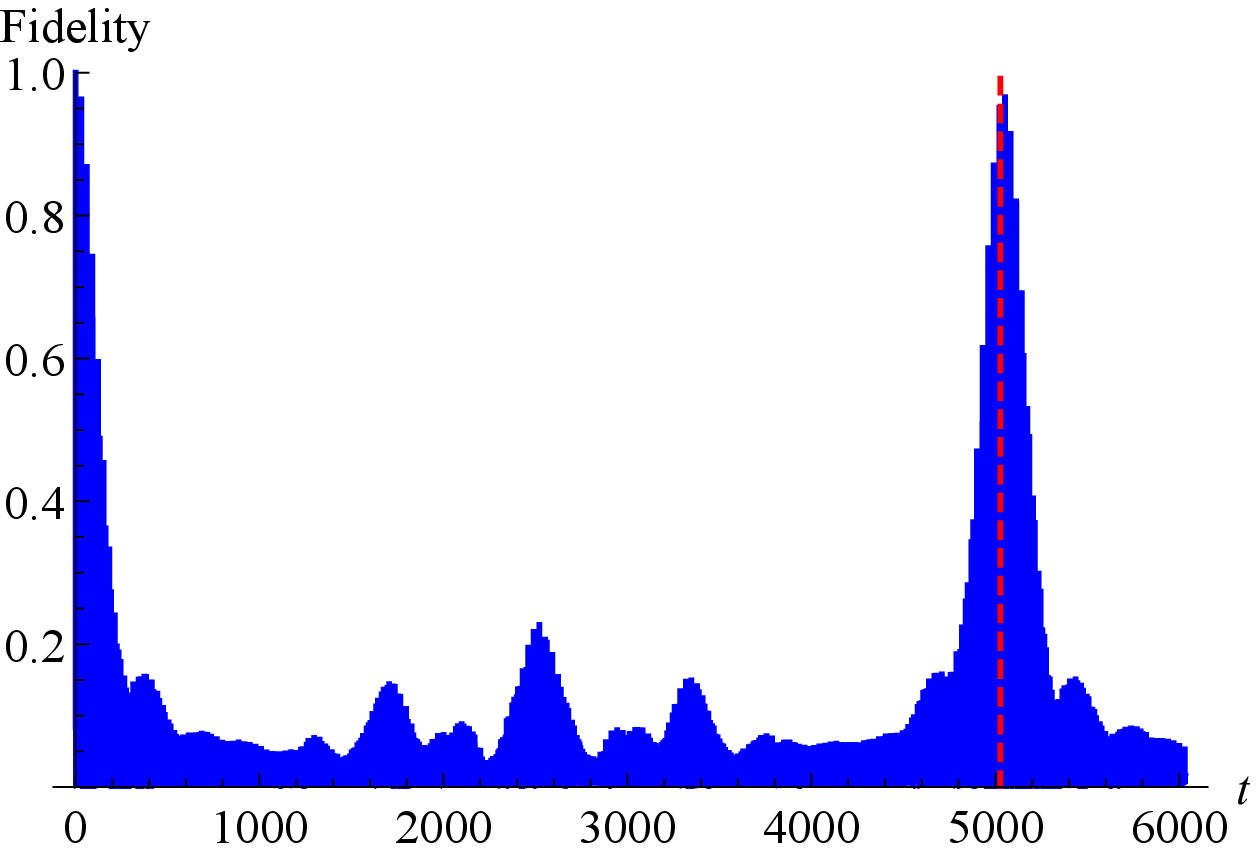}
\end{tabular}
\end{minipage}
\begin{minipage}{0.33\hsize}
\begin{tabular}{c}
(b) $N=11$\\
\includegraphics[width=5cm]{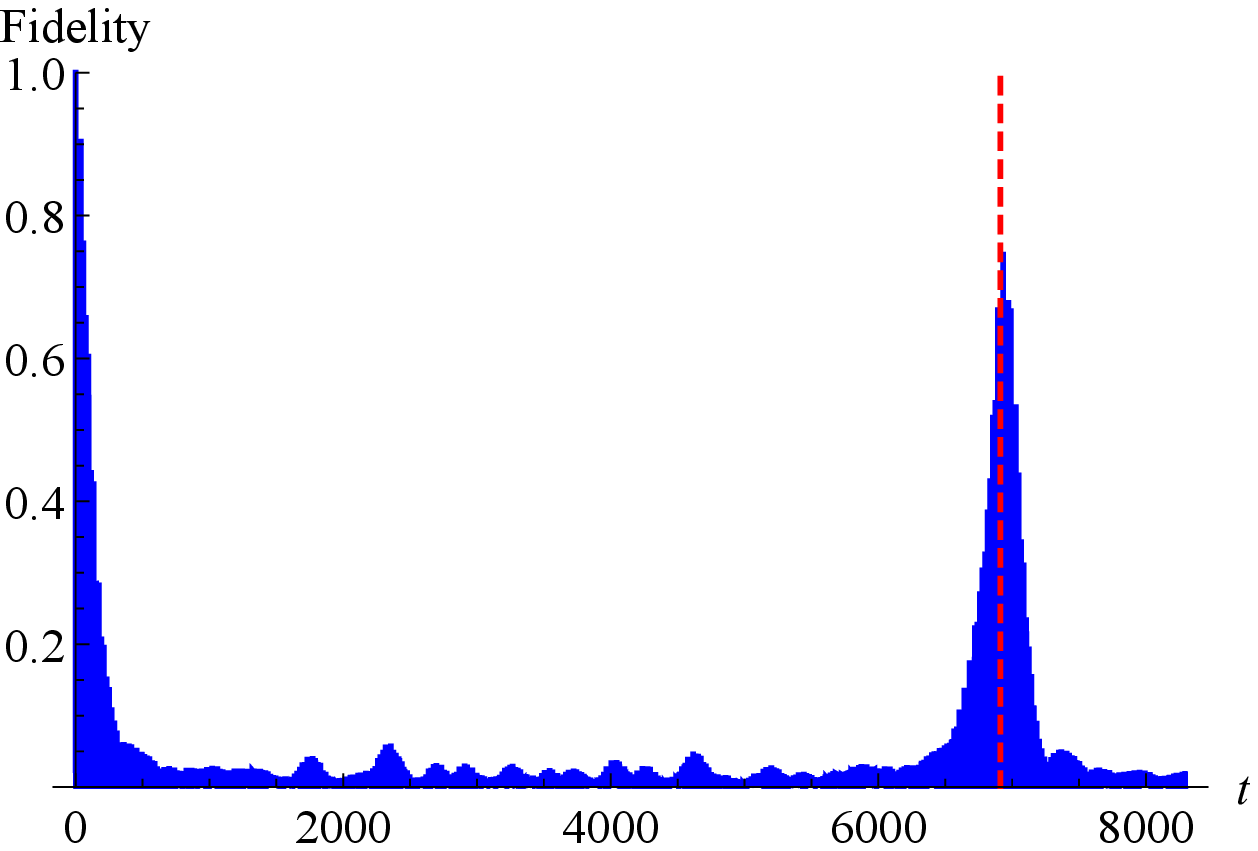}
\end{tabular}
\end{minipage}\begin{minipage}{0.33\hsize}
\begin{tabular}{c}
(c) $N=14$\\
\includegraphics[width=5cm]{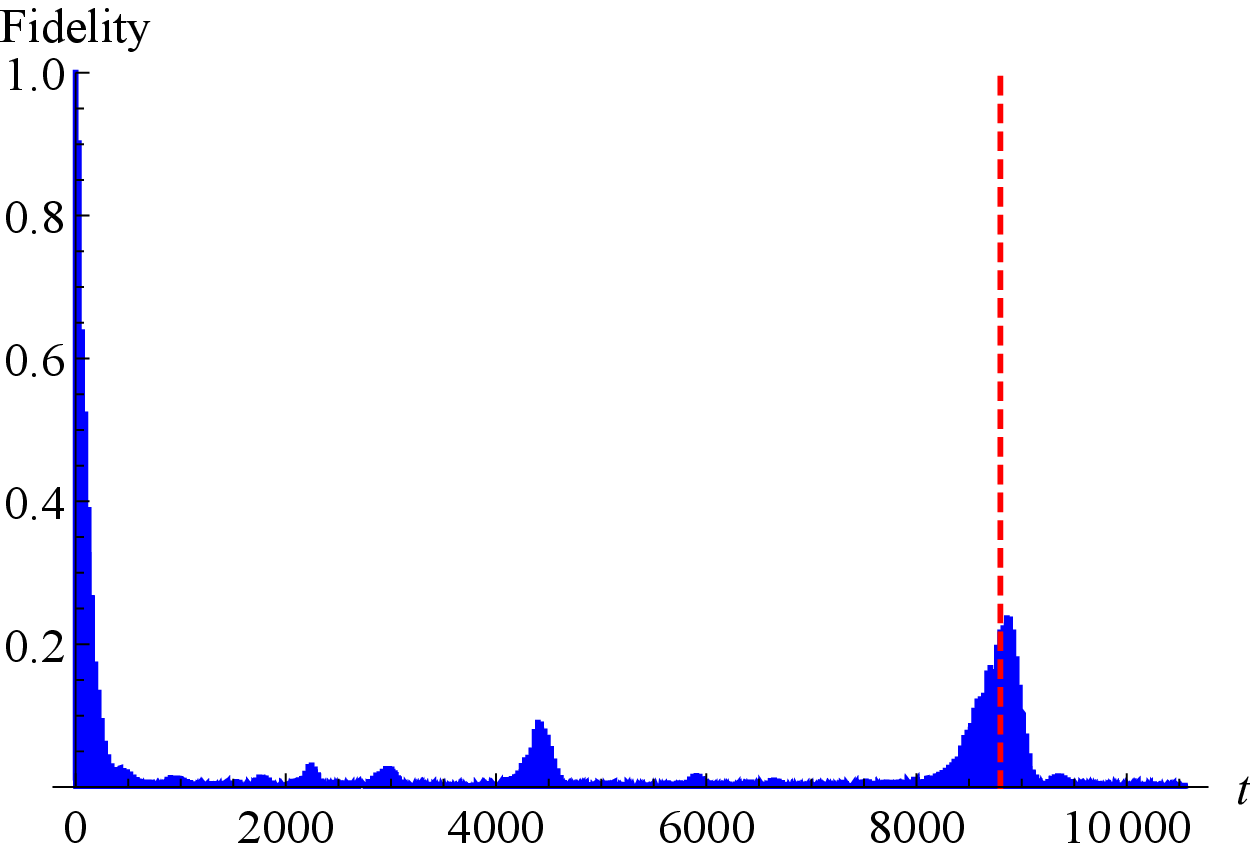}
\end{tabular}
\end{minipage}
\caption{The time evolution of the fidelity $\mathcal{F}(t)$.
The red dashed lines show the recurrence times predicted by eq.~(\ref{eq:rec_weak}).}
\label{fig:fidelity_weak}
\end{figure*}

First, we show numerical results for the weak-coupling regime.
We set $c=0.005$ and $\beta=0.5$.
We calculate the time evolution of the fidelity for several system sizes.
Figure~\ref{fig:fidelity_weak} shows numerical results for (a) $L=8$, (b) $L=11$, and (c) $L=14$.
We find that the recurrence actually occurs at the recurrence time predicted theoretically in eq.~(\ref{eq:rec_weak}), which is expressed by red dashed lines in Fig.~\ref{fig:fidelity_weak}.
As for the conditions~(\ref{eq:condition}) and (\ref{eq:condition2}), the values of $T_w^{(0)}\Delta V_w$ and $T_w\Delta E_w'$ are summarized in Table~\ref{table:weak}.
We see that $T_w^{(0)}\Delta V_w$ is very small up to $L=14$ and the condition~(\ref{eq:condition}) is satisfied, but $T_w\Delta E_w'$ becomes large as the system size increases.
As is seen in Fig.~\ref{fig:fidelity_weak}, the peak height of the fidelity near the predicted recurrence time decreases as the system size increases.
This is because the value of $T_w\Delta E_w'$ exceeds unity and the condition~(\ref{eq:condition2}) is not satisfied for large system sizes.

\begin{figure*}[tb]
\begin{tabular}{ccc}
\begin{minipage}{0.33\hsize}
\begin{tabular}{c}
(a) $N=10$\\
\includegraphics[width=5cm]{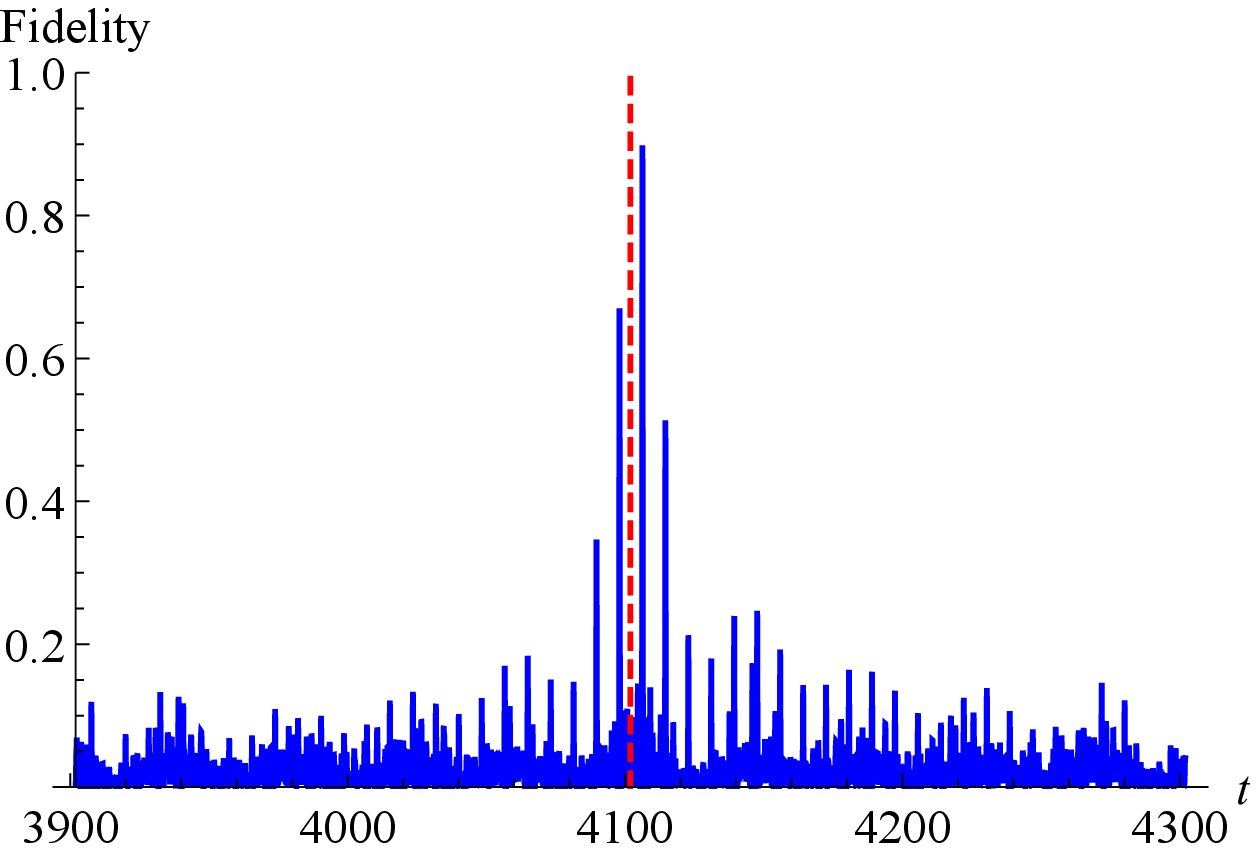}
\end{tabular}
\end{minipage}&
\begin{minipage}{0.33\hsize}
\begin{tabular}{c}
(b) $N=14$\\
\includegraphics[width=5cm]{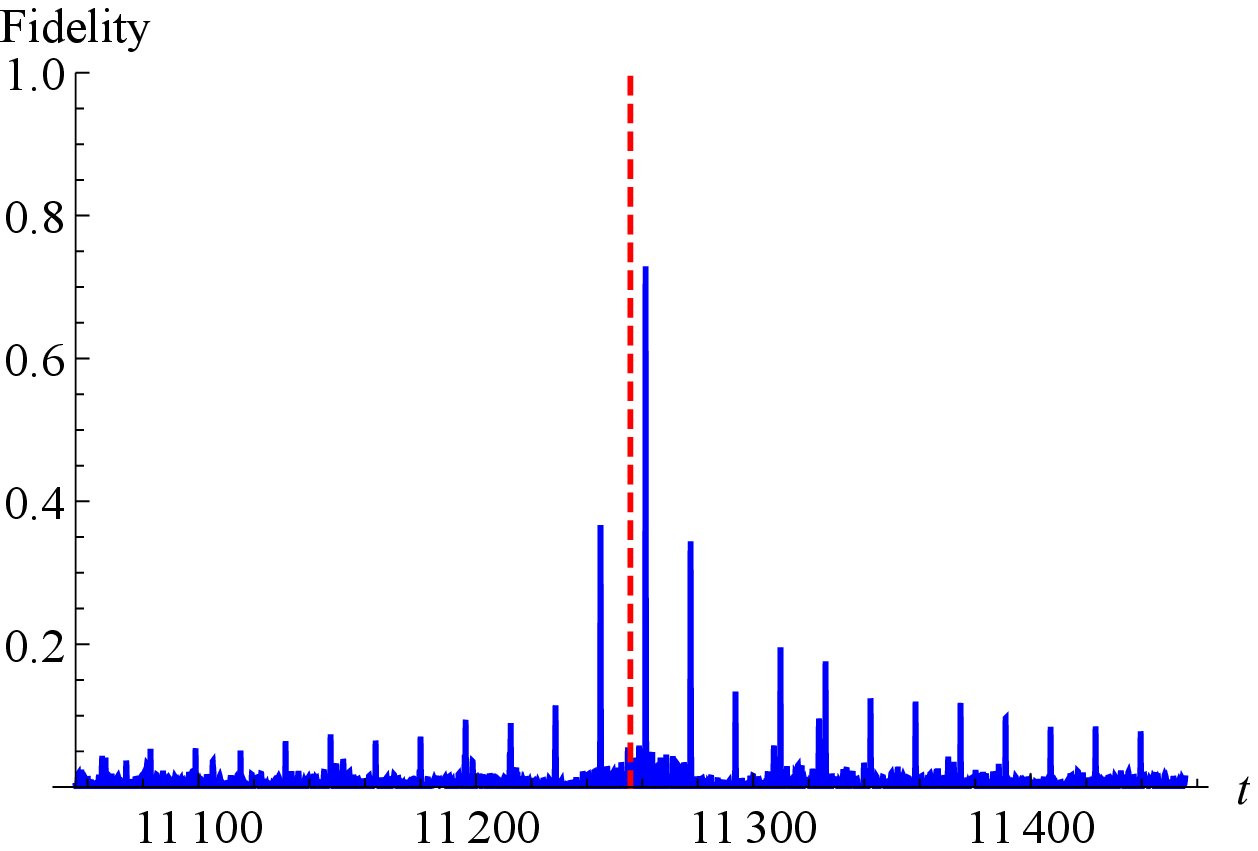}
\end{tabular}
\end{minipage}&
\begin{minipage}{0.33\hsize}
\begin{tabular}{c}
(c) $N=20$\\
\includegraphics[width=5cm]{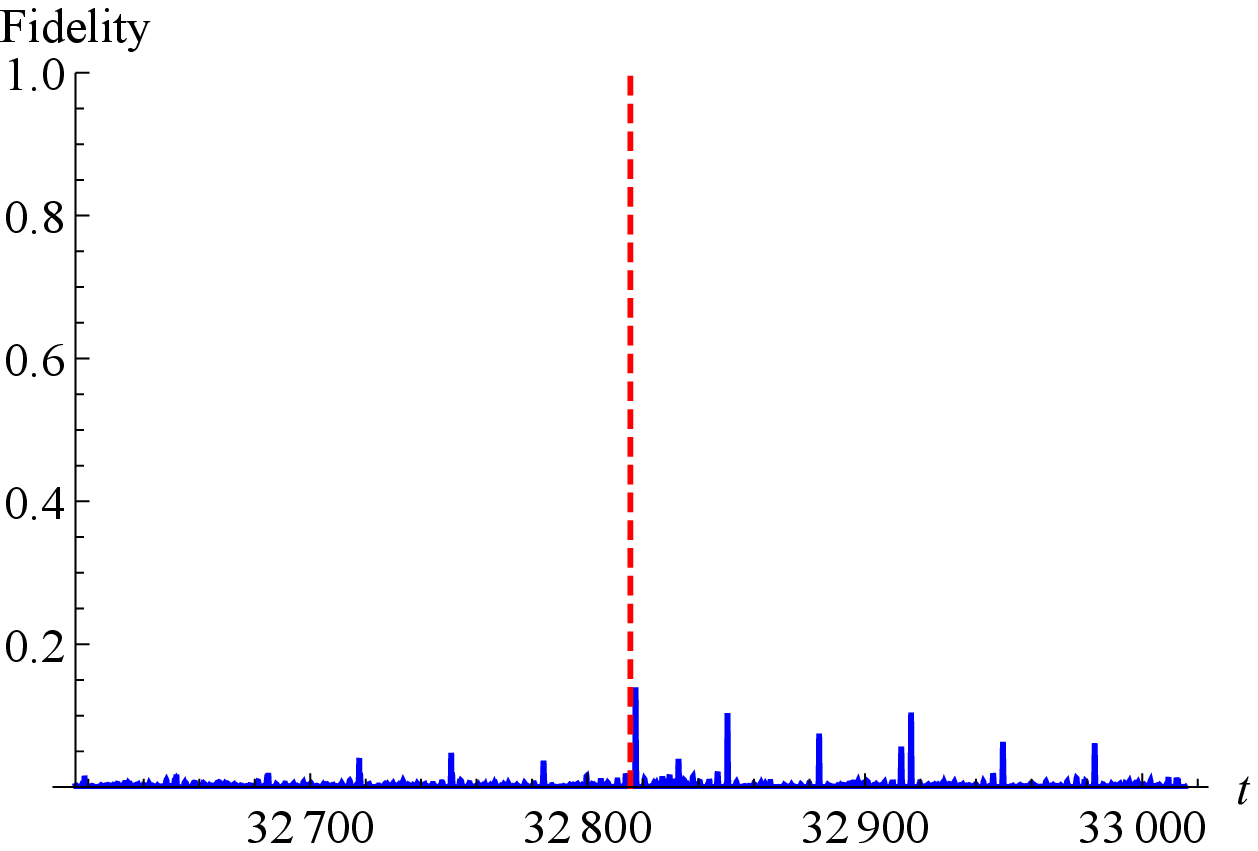}
\end{tabular}
\end{minipage}
\vspace{1cm}\\
\begin{minipage}{0.33\hsize}
\begin{tabular}{c}
(d) $N=11$\\
\includegraphics[width=5cm]{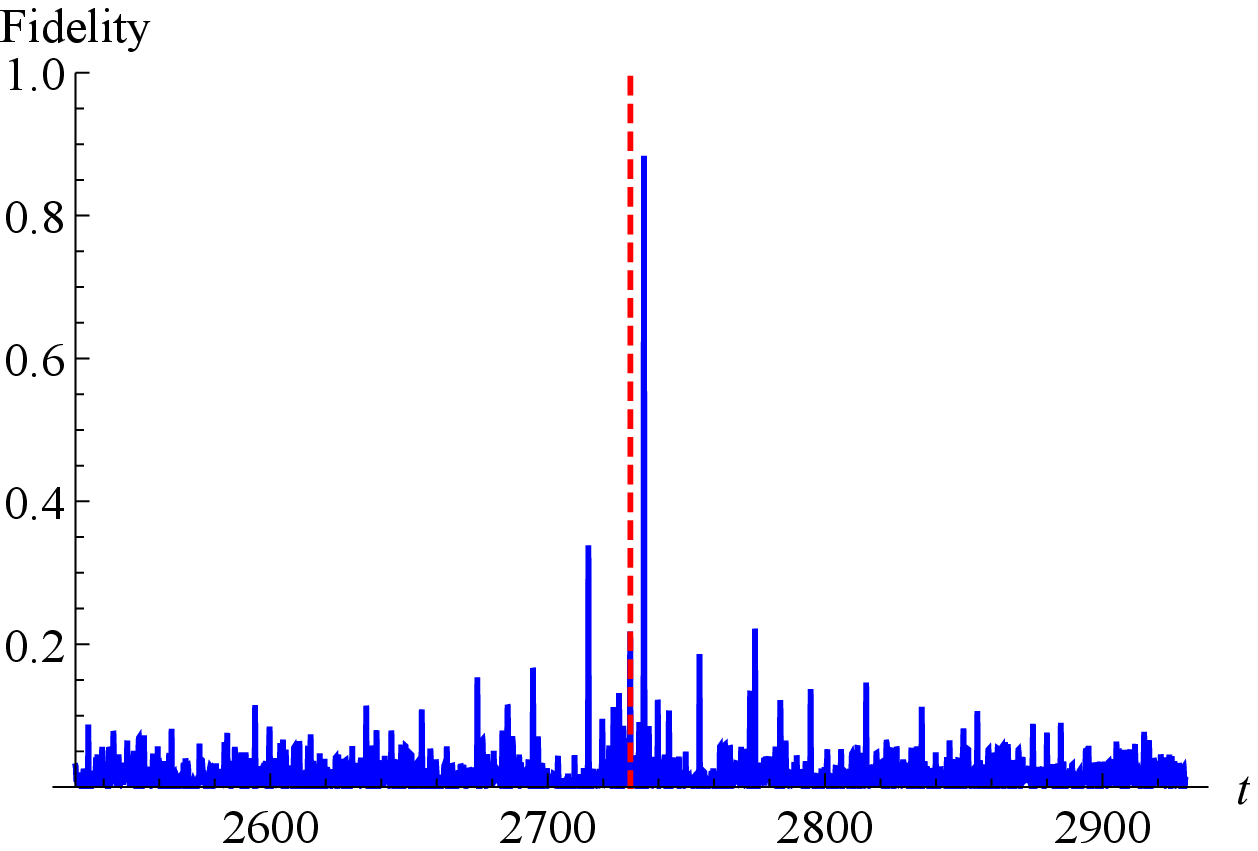}
\end{tabular}
\end{minipage}&
\begin{minipage}{0.33\hsize}
\begin{tabular}{c}
(e) $N=15$\\
\includegraphics[width=5cm]{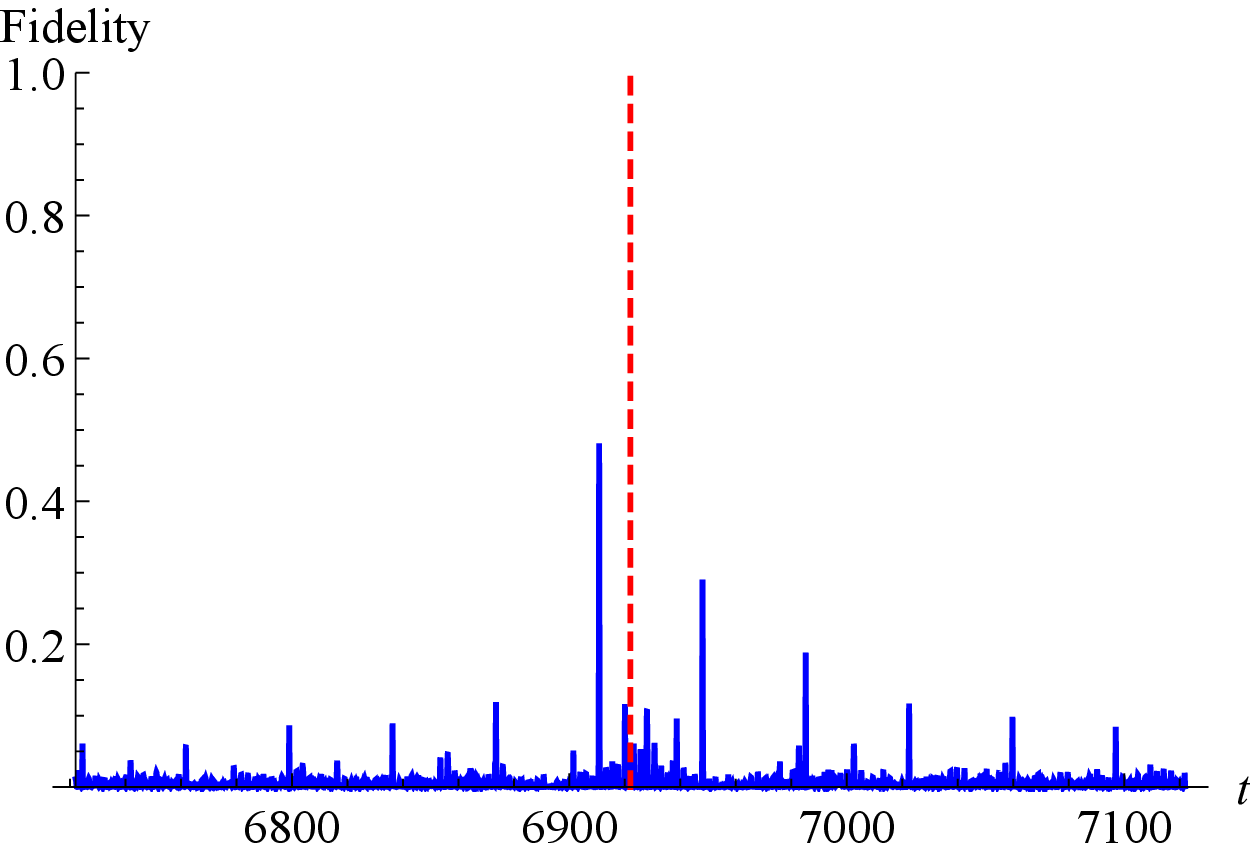}
\end{tabular}
\end{minipage}&
\begin{minipage}{0.33\hsize}
\begin{tabular}{c}
(f) $N=21$\\
\includegraphics[width=5cm]{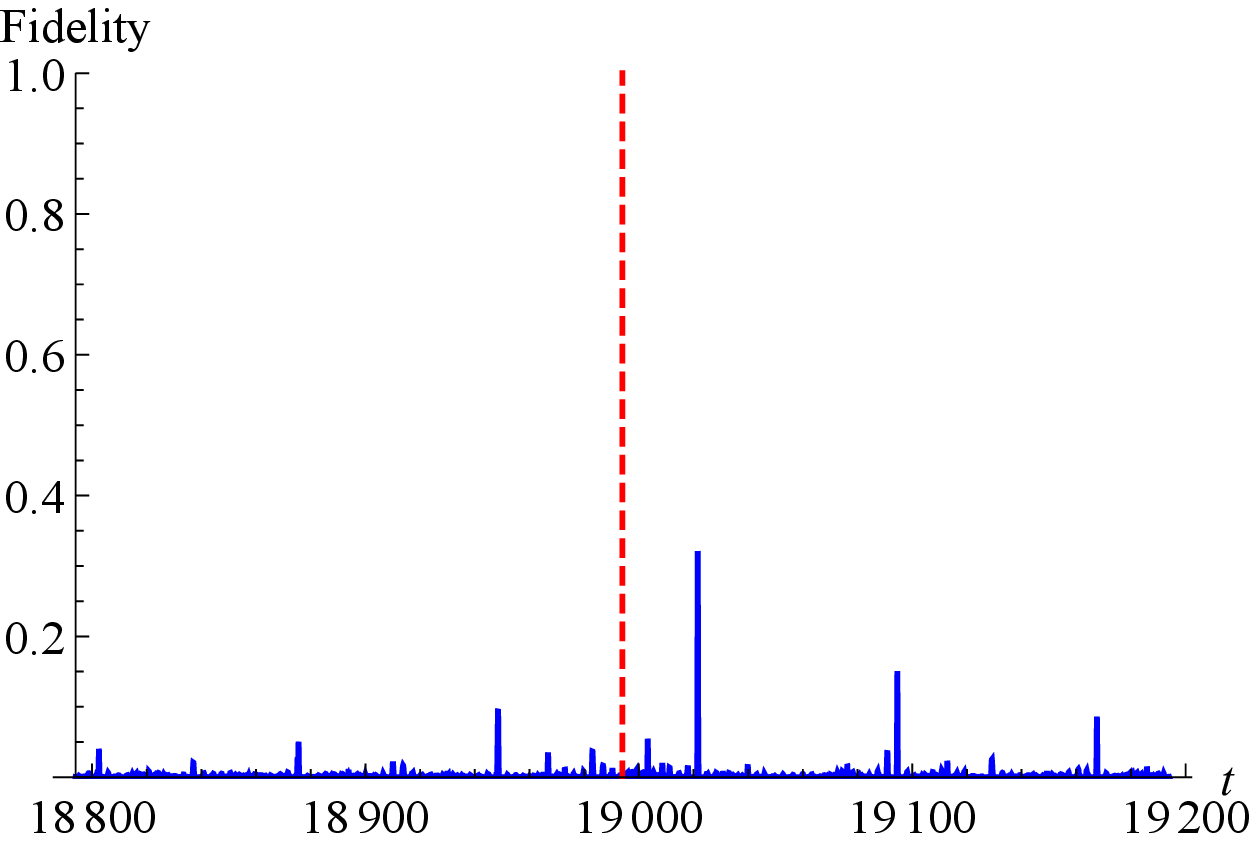}
\end{tabular}
\end{minipage}
\end{tabular}
\caption{The time evolution of the fidelity $\mathcal{F}(t)$ for $L=100$ and $\beta=0.2$.
The red dashed lines show the recurrence times predicted by eq.~(\ref{eq:rec_strong}).}
\label{fig:fidelity_strong}
\end{figure*}

Next, we show numerical results for the strong-coupling regime.
We set $c=100$ and $\beta=0.2$.
Figure~\ref{fig:fidelity_strong} shows numerical results of the time evolution of the fidelity for (a) $L=10$, (b) $L=14$, (c) $L=20$, (d) $L=11$, (e) $L=15$, and (f) $L=21$.
It is noted that the recurrence actually occurs at the recurrence time predicted in eq.~(\ref{eq:rec_strong}), which is expressed by red dashed lines in Fig.~\ref{fig:fidelity_strong}.
The values of $T_s^{(0)}\Delta V_s$ and $T_s\Delta E_s'$ are shown in Table~\ref{table:strong}.
We find that the peak height of the fidelity decreases as the system size increases.
This is because both $T_s^{(0)}\Delta V_s$ and $T_s\Delta E_s'$ increase as the system size increases.

\begin{table}
\begin{tabular}
{|l|c|c|c|}
\hline
$L$ &10 & 14 & 20\\
\hline\hline
$T_w^{(0)}\Delta V_w$ & 0.0121 & 0.0267 & 0.0490\\
$T_w\Delta E_w'$ & 0.221 & 0.690 & 1.62\\
\hline
\end{tabular}
\caption{The values of $T_w^{(0)}\Delta V_w$ and $T_w\Delta E_w'$ for each system size in the weak-coupling regime, $c=0.005$ and $\beta=0.5$.}
\label{table:weak}
\end{table}

\begin{table}
\begin{tabular}
{|l|c|c|c||c|c|c|}
\hline
$L$ &10 & 14 & 20 & 11 & 15 & 21\\
\hline\hline
$T_s^{(0)}\Delta V_s$ & 0.413 & 0.793 & 1.55 & 0.998 & 1.81 & 3.39\\
$T_s\Delta E_s'$ & 0.871 & 1.90 & 4.21 & 0.546 & 1.11 & 2.35\\
\hline
\end{tabular}
\caption{The values of $T_s^{(0)}\Delta V_s$ and $T_s\Delta E_s'$ for each system size in the strong-coupling regime, $c=100$ and $\beta=0.2$.}
\label{table:strong}
\end{table}

\section{Recurrence time for an initial state of a superposition of yrast states}
\label{sec:dark}

The recurrence times $T_w$ and $T_s$ in Eqs.~(\ref{eq:rec_weak}) and (\ref{eq:rec_strong}), respectively, are valid for generic initial states, but some initial states accidentally exhibit shorter recurrence times.
As such an example, we consider the initial state given by a uniform superposition of yrast states,
\beq
|\Psi(0)\>=\frac{1}{\sqrt{N}}\sum_{K=1}^Ne^{2\pi iKx/L}|K\>,
\label{eq:dark}
\eeq
where $|K\>$ is the yrast state of the momentum $2\pi K/L$.
An yrast state is an energy eigenstate with the lowest energy for a given total momentum and correspond to Lieb's type-II excitations~\cite{Lieb1963}.
The Bethe quantum numbers for the yrast state $\ket{K}$ are given by
\beq
I_j=\left\{
\begin{split}
&j-\frac{N+1}{2} \qquad \text{for $1\leq j\leq N-K$}, \\
&j-\frac{N-1}{2} \qquad \text{for $N-K+1\leq j\leq N$}.
\end{split}
\right.
\eeq
Considering this initial state is particularly interesting because it has been argued that the state~(\ref{eq:dark}) is considered to represent the quantum counterpart of the dark soliton solution in the classical nonlinear Schr\"odinger equation~\cite{Sato2012,Sato2016,Kaminishi_arXiv2018}.

\subsection{Weak-coupling case}
In the weak-coupling regime, the state $|K\>$ corresponds to the case where $a=1$ or $2$, $n_1=N-K$, $n_2=K$, $k^{(0)}(1)=0$, and $k^{(0)}(2)=2\pi/L$.
In this case,
\beq
E_w^{(0)}(\bm{k}^N)=\left(\frac{2\pi}{L}\right)^2K,
\label{eq:E0_w_yrast}
\eeq
and
\beq
V_w(\bm{k}^N)=\frac{2c}{L}K(N-K).
\label{eq:V_w_yrast}
\eeq

When $N$ is odd, $K(N-K)$ is even for any integer $K$.
Therefore, the recurrece time is given by $T\simeq \pi L/2c$.

On the other hand, when $N$ is even, $K(N-K)$ is odd for odd $K$ and even for even $K$, and hence one is tempted to consider that the recurrence time is $T\simeq \pi L/c$ for even $N$.
However, it is not the case.
The recurrence time is also given by $T\simeq \pi L/2c$ for even $N$.
Actually, at $T=\pi L/2c$,
\beq
V_w(\bm{k}^N)T=\theta_K+2\pi m_K',
\label{eq:VT_yrast}
\eeq
where $m_K'$ is integer and $\theta_K=0$ for even $K$ and $\theta_K=\pi$ for odd $K$.
While, when $T\simeq T_w^{(0)}/2=(L^2/4\pi)\times$(odd integer), $E_w^{(0)}(\bm{k}^N)T$ is given by
\beq
E_w^{(0)}(\bm{k}^N)T\simeq\theta_K+2\pi m_K,
\label{eq:ET_yrast}
\eeq
where $m_K$ is integer.
By adding eq.~(\ref{eq:VT_yrast}) and (\ref{eq:ET_yrast}), we obtain
\beq
E_w^{(0)}(\bm{k}^N)T+V_w(\bm{k}^N)T\simeq 2\pi m_K'',
\eeq
where $m_K''$ is some integer.
The recurrence time for the initial state (\ref{eq:dark}) is thus given by
\beq
T=\frac{\pi L}{2c}
\eeq
both for even $N$ and for odd $N$.
This recurrence time is half of that in the general case~(\ref{eq:rec_weak}), which is a special case not satisfying eqs.~(\ref{eq:rec0}) and (\ref{eq:rec1}).

\subsection{Strong-coupling case}
In the strong-coupling regime, the state $|K\>$ corresponds to the case where $I_{\rm tot}^{(1)}=K$ and $I_{\rm tot}^{(2)}=K(N+1-K)+{\rm const.}$, and thus
\beq
V_s(\bm{k}^N)=-\frac{16\pi^2}{L^3c}\left(1-\frac{\rho}{3c}\right)(N+1)K(N-K)+{\rm const}.
\eeq
Here, $K(N-K)$ is always even for odd $N$, which halves the recurrence time for odd $N$.
The recurrence time for the initial state~(\ref{eq:dark}) in the strong coupling regime is therefore
\beq
T\simeq
\left\{
\begin{split}
\frac{L^2c}{16\pi\rho\left(1-\frac{\rho}{3c}\right)} \qquad \text{for odd $N$}, \\
\frac{L^2c}{8\pi\rho\left(1-\frac{\rho}{3c}\right)} \qquad \text{for even $N$}.
\end{split}
\right.
\eeq
Thus the recurrence time grows as the square of the system size with $\rho$ held fixed, which is consistent with the numerical finding in Ref.~\cite{Kaminishi2015}.
This system-size dependence is different from that for generic case given by Eq.~(\ref{eq:rec_strong}).

\section{Conclution}
\label{sec:conclusion}

We have evaluated the recurrence time in the one-dimensional Bose gas with the delta function interactions via perturbative analyses of the Bethe ansatz equation.
We have shown that the recurrence time grows only polynomially in $N$ for any initial state in the weak- and strong-coupling regimes as long as this initial state satisfies certain conditions, i.e., eqs.~(\ref{eq:condition}) and (\ref{eq:condition2}) for the weak-coupling regime, and eqs.~(\ref{eq:condition_s}) and (\ref{eq:condition2_s}) for the strong-coupling regime.
Since these conditions are not fulfilled for sufficiently large $N$, short recurrence times predicted in this work is particularly relevant for small systems.
Although our analysis is on a concrete integrable model, i.e., the Lieb-Liniger model, it is a future problem to understand whether this kind of short recurrence times in small quantum systems is universal or not.

\begin{acknowledgments}
We thank T. Deguchi and J. Sato for useful discussions.
The present work was supported by JSPS KAKENHI Grants No. JP16J03140 and Grants-in-Aid for Scientific Research C (No. 18K03444) from MEXT of Japan.
\end{acknowledgments}

\bibliography{RT_2018}

\end{document}